\documentclass[aps,pra,twocolumn,showpacs,superscriptaddress]{revtex4-2}
\usepackage{graphicx}
\usepackage{dcolumn}
\usepackage{bm}
\usepackage{color}
\usepackage{times}
\usepackage{amssymb}
\usepackage{amsmath}
\usepackage{epsfig}
\usepackage{epstopdf}
\usepackage{dsfont}
\usepackage{txfonts}
\usepackage{subfigure}
\usepackage[colorlinks=true,citecolor=blue, linkcolor=blue,urlcolor=blue]{hyperref}
\usepackage{setspace}
\usepackage{bookmark}
\usepackage{color}

\newenvironment{proof}{{\indent  \it Proof:\,}}{\hfill $\blacksquare$\par}
\definecolor{Green3}{rgb}{0.80,0.87,0.76}

\begin{document}

\title{Maximum residual strong  monogamy inequality for multiqubit entanglement}

\author{Dong-Dong Dong}
 \affiliation{School of Physics and Optoelectronic Engineering, Anhui University, Hefei
230601,  People's Republic of China}
\author{Xue-Ke Song}%
 \affiliation{School of Physics and Optoelectronic Engineering, Anhui University, Hefei
230601,  People's Republic of China}

 \author{Liu Ye}
  \affiliation{School of Physics and Optoelectronic Engineering, Anhui University, Hefei
230601,  People's Republic of China}
\author{Dong Wang}
 \email{dwang@ahu.edu.cn}
  \affiliation{School of Physics and Optoelectronic Engineering, Anhui University, Hefei
230601,  People's Republic of China}

\author{Gerardo Adesso}
\email{gerardo.adesso@nottingham.ac.uk}
  \affiliation{School of Mathematical Sciences and Centre for the Mathematical and Theoretical Physics of Quantum Non-Equilibrium Systems,
 University of Nottingham, University Park, Nottingham NG7 2RD, United Kingdom}

\date{\today}

\begin{abstract}
 {We establish two new inequalities, the weighted strong monogamy (WSM) and the maximum residual  strong monogamy (MRSM), }
which sharpen the generalized Coffman-Kundu-Wootters inequity for multiqubit states.
The WSM inequality distinguishes itself from the strong monogamy (SM) conjecture [\href{https://journals.aps.org/prl/abstract/10.1103/PhysRevLett.113.110501}{Phys. Rev. Lett. \textbf{113}, 110501 (2014)}] by using coefficients rather than exponents to modulate the weight allocated to various \emph{m}-partite contributions.
 In contrast, the MRSM inequality is formulated using only the maximum $m$-partite entanglement.
We find that the residual entanglement of the MRSM inequality can effectively distinguish the separable states.
We also compare the tightness of various SM inequalities and provide {examples using a four-qubit mixed state and a five-qubit pure state}  to illustrate the MRSM inequality.
{These examples characterize the trade-off relations among entanglement components involving varying numbers of qubits.}
 Our results provide a rigorous framework to characterize and quantify the monogamy of multipartite entanglement.
\end{abstract}

\maketitle

 \section{Introduction}\label{sec:intro}
Quantum entanglement, one of the most intriguing features of quantum mechanics, captures the fundamental distinction between the quantum and classical worlds. It serves as an indispensable resource in modern quantum technologies, underpinning applications across quantum information processing, while also playing a pivotal role in diverse domains such as condensed matter physics, statistical mechanics, and thermodynamics \cite{RevModPhys.81.865,RevModPhys.80.517}.
While bipartite entanglement has been extensively studied over recent decades \cite{PhysRevA.54.3824,PhysRevLett.78.5022,PhysRevLett.80.2245,PhysRevA.64.042315,PhysRevA.65.032314,PhysRevA.106.042415,PhysRevA.107.052403}, the quantification of multipartite entanglement --- often essential for quantum information tasks --- remains a profound challenge. A crucial aspect of this challenge is understanding how entanglement is distributed among the constituents of a multipartite system, which is a central goal of this work.

The fact that entanglement cannot be freely shared in multipartite quantum systems, known as \emph{monogamy}, is a phenomenon with no classical counterpart.
  In 2000, Coffman \emph{et al.} firstly formalized the monogamy of entanglement for a three-qubit system via an inequality, known as Coffman-Kundu-Wootters (CKW) inequality \cite{PhysRevA.61.052306}
\begin{align}
\tau_{A(BC)} \geq \tau_{AB} + \tau_{AC},
  \label{Eq.1}
    \end{align}
where $\tau_{A(BC)}$ represents the square of  bipartite concurrence for the bipartition $A:BC$. Additionally, $\tau_{AB}$ and $\tau_{AC}$ denote squared concurrence of reduced states $\rho_{AB}$ and $\rho_{AC}$, respectively.
Notably, for pure states, the difference between the left- and right-hand side of Eq.~(\ref{Eq.1}) can serve  as a measure of residual tripartite entanglement, also known as three-tangle
\begin{align}
\tau_{ABC}:=\tau_{A(BC)} - \tau_{AB} - \tau_{AC},
  \label{Eq.2}
    \end{align}
which is invariant under qubit permutation.
In 2006, the generalized version of the CKW inequality  was proven by Osborne and Verstraete \cite{PhysRevLett.96.220503},
\begin{align}
\tau_{q_1(q_2...q_n)} \geq \tau_{q_1q_2} + \tau_{q_1q_3}+...+\tau_{q_1q_n}.
  \label{Eq.3}
    \end{align}
Following that, similar monogamy inequalities were established for other entanglement measures, including the squared negativity \cite{PhysRevA.75.062308} and the squared  entanglement of formation \cite{PhysRevA.89.034303,PhysRevLett.113.100503}. In addition, the minimum exponents required for monogamy inequalities were investigated \cite{PhysRevA.90.024304}.
In 2014, a strong monogamy (SM) conjecture for multiqubit entanglement was proposed by Regula \emph{et al.} \cite{PhysRevLett.113.110501}, involding additional terms related to $m$-partite entanglement on the right side of Eq.~(\ref{Eq.3}), where $m$ ranges from $3$ to $n-1$. The SM inequality takes the form:
\begin{align}
\tau_{q_1(q_2\cdots q_n)} \geq \sum_{j=2}^n\tau_{q_1q_j} +\sum_{m=3}^{n-1} \sum_{\vec{j}^m : S_n} [\tau_{q_1q_{j^m_1}... q_{j^m_{m-1}}}]^{\mu_m},
\label{Eq.4}
\end{align}
where $\vec{j}^m : S_n$ implies that  the index vector $\vec{j}^m=(j_1^m,...,j_{m-1}^m)$ runs through all possible combinations of $m-1$ distinct elements from the index set $S_n=\{2,...,n \}$. The rational exponents $\{\mu_m\ |\ \mu_m \ge 1\}$ are introduced to regulate the weight assigned to different $m$-partite contributions. Note that $\mu_m $ should be as small as possible, since the verification of Eq.~(\ref{Eq.4}) for a specific set  $\{\mu_m^\star\}$ implies its validity for all $\{\mu_m\} \succeq \{\mu_m^\star\}$.
In addition, $\tau_{q_1q_{j^m_1}... q_{j^m_{m-1}}}$ can be regarded as the residual entanglement of $m$-partite reduced states, obtained via  a  convex roof construction of the residual entanglement for pure states, which is defined recursively as the difference between two sides of Eq.~(\ref{Eq.4}). Subsequently, other SM conjectures were proposed for the square of convex roof extended negativity  \cite{PhysRevA.92.042307} and the squared negativity \cite{PhysRevA.93.012327}. However, these promising conjectures have yet to be rigorously proven. Furthermore, the minimal choice of exponent $\mu_m=1$, which is appropriate for continuous variable Gaussian states \cite{GaussianSM}, was found inadequate even for four-qubit states \cite{PhysRevA.93.052338}. While it has been numerically shown that  $\mu_m=m/2$ is  valid for four-qubit states \cite{PhysRevLett.113.110501}, the suitable $\mu_m$ for systems with a larger number of qubits remains unknown.

In this article,  we present different general approaches to quantify the monogamy of multipartite entanglement.
We first utilize coefficients rather than exponents to modulate the weight allocated to various $m$-partite contributions, thereby analytically deriving a weighted  SM (WSM) inequality. We then establish the maximum residual SM (MRSM) inequality, which considers only the maximum entanglement of
$m$-partite contributions, in contrast to conventional methods that sum over all such terms for continuous-variable systems \cite{GaussianSM}.  We prove the inequalities by using squared concurrence for an arbitrary number of qubits, and we demonstrate that they can be naturally extended to other entanglement measures (Section~\ref{sec:ineq}).
Furthermore, we compare the tightness of different SM inequalities (Section~\ref{sec:compare}). Finally, we present examples using a four-qubit mixed state (Section~\ref{sec:fourmix}) and a five-qubit pure state (Section~\ref{sec:fivepure}) to illustrate the MRSM inequality.

 \section{Defining the WSM and MRSM Inequalities}\label{sec:ineq} 
 The squared bipartite concurrence, also known as the tangle (or one-tangle), for any pure state $|\psi \rangle_{AB}$ of a $2 \otimes d$ system is given by \cite{PhysRevA.64.042315}
\begin{align}
\tau_{AB}(|\psi\rangle_{AB})=2[1-{\rm Tr}(\rho_A^2)]=4{\rm det}(\rho_A),
\label{Eq.5}
\end{align}
where $\rho_A$ is the reduced matrix of the subsystem $A$. The tangle for any mixed state $\rho_{AB}$ is defined via the convex roof construction
\begin{align}
\tau_{AB}(\rho_{AB}):=\inf_{\{p_i, |\psi_i\rangle\}} \sum_i p_i \mathcal{\tau}_{AB}(|\psi_i \rangle),
\label{Eq.6}
\end{align}
where the infimum runs over all pure-state decompositions of $\rho_{AB}$, $\rho_{AB}=\sum_i p_i |\psi_i\rangle \langle \psi_i|$.
 Then for arbitrary states  of a $2 \otimes 2 \otimes 2^{n-2}$ system $ABC$, the CKW monogamy inequality holds \cite{PhysRevLett.96.220503}:
 \begin{align}
\tau_{A(BC)} \geq \tau_{AB} + \tau_{AC}.
\label{Eq.7}
\end{align}

We will now establish the first result of this paper.

{\it Theorem 1.\,} For arbitrary $n$-qubit states, the WSM inequality holds, which can be expressed as
\begin{align}
\tau_{q_1(q_2\cdots q_n)} \! \geq \! \sum_{j=2}^n\tau_{q_1q_j} \! +\! \sum_{m=3}^{n-1}\binom{n-1}{m-1} ^{-1} \! \sum_{\vec{j}^m:S_n} \tau_{q_1q_{j^m_1}... q_{j^m_{m-1}}},
\label{Eq.8}
\end{align}
where $\binom{n-1}{m-1}=\frac{(n-1)!}{(m-1)!(n-m)!} $ is the binomial coefficient. The index vector $\vec{j}^m:S_n$  is defined as in Eq.~(\ref{Eq.4}).

 {The proof is offered in  } \hyperlink{app1}{Appendix}.
Note that for each \emph{m}, the expression $\binom{n-1}{m-1} ^{-1} \! \sum_{\vec{j}^m:S_n} \tau_{q_1q_{j^m_1}... q_{j^m_{m-1}}}$ in fact represents the average of the distinct $m$-partite ``residual entanglements'' whose index vector is  $\vec{j}^m$.
From the derivation, we observe that the WSM inequality can be naturally extended to other entanglement measures, provided that they satisfy the following conditions:
 (a) For arbitrary states of a $2\otimes 2\otimes 2^{n-2}$ system \emph{ABC},  the measure satisfies an inequality of the form Eq.~(\ref{Eq.1});
 (b) the measure is convex.
 For example, the  entanglement of formation to the $\sqrt{2}$ power  satisfies conditions (a) \cite{LUO2015511} and (b) \cite{PhysRevA.54.3824}.

  {Unfortunately, the difference between left- and right-hand sides of the WSM inequality  cannot be  considered as a multipartite entanglement measure, since it can be positive for certain separable states.
 Specifically, for four-qubit states, the WSM inequality takes the form of
   \begin{align}
\tau_{A(BCD)} \! \geq \! \tau_{AB} \!+\! \tau_{AC} \!+\! \tau_{AD} \!+\! \frac{1}{3} ( {\cal \tau}_{ABC}\!+\!{\cal \tau}_{ABD}\!+\!{\cal \tau}_{ACD}).
\label{Eq.b9}
\end{align}
Then for the state $|\psi \rangle_{ABCD} =1/\sqrt{2} \left( |000\rangle +|111\rangle \right) \otimes |0\rangle$, the difference between left- and right-hand sides of the WSM inequality evaluates to 2/3. To address this issue, we introduce a new monogamy inequality.
}

  {
{\it Theorem 2.\,} For arbitrary $n$-qubit states, the following inequality holds
\begin{align}
\tau_{q_1(q_2\cdots q_n)} \! \geq \! \sum_{j=2}^n\tau_{q_1q_j} \! +\! \max \limits_{\vec{j}^m:S_{m+1}}\left\{ \sum_{m=3}^{n-1}\!  \tau_{q_1q_{j^m_1}... q_{j^m_{m-1}}} \right\},
\label{Eq.b10}
\end{align}
where  $\vec{j}^m : S_{m+1}$ implies that  the index vector $\vec{j}^m=(j_1^m,...,j_{m-1}^m)$ runs through all possible combinations of $m-1$ distinct elements from the index set $S_{m+1}$. The set $S_{m+1}$  is defined by the indexes  of $\vec{j}^{m+1}$, and the original set $S_n=\{2,...,n\}$.
The $m$-partite residual entanglement $\tau_{q_1q_{j^m_1}... q_{j^m_{m-1}}}$ is obtained via a convex roof construction of the residual entanglement for pure states, which is formulated as the difference between left- and right-hand sides of Eq.~(\ref{Eq.b10}).
Specifically, the residual entanglement for any $n$-qubit pure state  is given by
\begin{align}
\tau_{q_1q_2...q_n} :=& \tau_{q_1(q_2\cdots q_n)} \! - \! \sum_{j=2}^n\tau_{q_1q_j} \!
-\!  \max \limits_{\vec{j}^m:S_{m+1}}\left\{ \sum_{m=3}^{n-1}\! \tau_{q_1q_{j^m_1}... q_{j^m_{m-1}}} \right\},
\label{Eq.b11}
\end{align}
and the residual entanglement for an arbitrary mixed state $\rho$ is expressed as
\begin{align}
\tau_{q_1q_2...q_n}:=\inf_{\{p_i, |\psi_i\rangle\}} \sum_i p_i \mathcal{\tau}_{q_1q_2...q_n}(|\psi_i \rangle),
\label{Eq.bb12}
\end{align}
where the infimum runs over all pure-state decompositions of $\rho=\sum_i p_i |\psi_i\rangle \langle \psi_i|$. Then the $m$-partite residual entanglement $\tau_{q_1q_{j^m_1}... q_{j^m_{m-1}}}$ is generated via Eqs.~(\ref{Eq.b11}) and (\ref{Eq.bb12})   for corresponding reduced states $\rho_{q_1q_{j^m_1}... q_{j^m_{m-1}}}$ in a recursive manner.
}

 {
\begin{proof}
Without loss of generality, for simplicity,  consider the maximum  index vector $\vec{j}^m=(2,...,m)$. Eq.~(\ref{Eq.b10}) reduces to
\begin{align}
\tau_{q_1(q_2\cdots q_n)} \! \geq \! \sum_{j=2}^n\tau_{q_1q_j} \! +\!  \sum_{m=3}^{n-1}\! \tau_{q_1...q_m}.
\label{Eq.b12}
\end{align}
Consider the pure states $|\psi_i\rangle$ belonging to an optimal decomposition of $\rho_{q_1(q_2\cdots q_{n-1})}$, that is, a decomposition that minimizes $\langle \tau_{q_1(q_2\cdots q_{n-1})} \rangle$, i.e., $\langle \tau_{q_1(q_2\cdots q_{n-1})} \rangle=\sum_i p_i \tau_{q_1(q_2\cdots q_{n-1})}(|\psi_i\rangle)$. Then
\begin{align}
\tau_{q_1q_2\cdots q_{n-1}} \! &\le \langle \tau_{q_1(q_2\cdots q_{n-1})} - \! \sum_{j=2}^{n-1}\tau_{q_1q_j} \! -\!  \sum_{m=3}^{n-2}\! \tau_{q_1...q_m} \rangle \nonumber \\
&=\langle \tau_{q_1(q_2\cdots q_{n-1})} \rangle -   \! \sum_{j=2}^{n-1} \langle \tau_{q_1q_j} \rangle \! -\!  \sum_{m=3}^{n-2}\! \langle \tau_{q_1...q_m} \rangle \nonumber \\
&\le \tau_{q_1(q_2\cdots q_{n-1})} - \! \sum_{j=2}^{n-1}\tau_{q_1q_j} \! -\!  \sum_{m=3}^{n-2}\! \tau_{q_1...q_m},
\label{Eq.b13}
\end{align}
with the first inequality due to  the convexity of $\tau_{q_1q_2\cdots q_{n-1}}$. The last inequality follows from the fact that for mixed states, any $\tau_{q_1q_j}$ is a convex function  of the density matrices \cite{PhysRevLett.80.2245}, and $\tau_{q_1...q_m}$ is defined via a convex roof construction.
From Eq.~(\ref{Eq.b13}), we have
\begin{align}
\begin{split}
\sum_{j=2}^n\tau_{q_1q_j} \! +\!  \sum_{m=3}^{n-1}\! \tau_{q_1...q_m} &\le \tau_{q_1q_n} \! +\!  \tau_{q_1(q_2\cdots q_{n-1})} \\
&\le \tau_{q_1(q_2\cdots q_n)} ,
\label{Eq.bb15}
\end{split}
\end{align}
in which the second inequality  is proven true according to Eq.~(\ref{Eq.7}).
When the maximum index vector $\vec{j}^m$  takes on other values, the term $\tau_{q_1q_n}  +  \tau_{q_1(q_2\cdots q_{n-1})}$ in  Eq.~(\ref{Eq.bb15}) corresponds to a different combination of the form $\tau_{q_1q_i}  +  \tau_{q_1(q_2\cdots q_{i-1}q_{i+1}...q_{n})}$ with $i\in \{ 2...n\}$. Nevertheless, Eq.~(\ref{Eq.bb15}) remains valid, as can be derived from Eq.~(\ref{Eq.7}).
Thereby,  the proof is completed.
\end{proof}
}

\smallskip

 {
For convenience, we call Eq.~(\ref{Eq.b10}) the {\em maximum residual strong  monogamy} (MRSM) inequality.
It is worth noting that for the fundamental case of a three-qubit system, the higher-order terms (where $m$ ranges from 3 to $n-1$) in both the WSM and MRSM inequalities vanish. Consequently, both inequalities naturally reduce to the standard CKW inequality Eq.~(\ref{Eq.7}), serving as a consistency check with established tripartite monogamy relations.
The derivation shows that the MRSM inequality can be extended to other entanglement measures under the same conditions as the WSM inequality.
 {Remarkably, the residual entanglement $\tau_{q_1q_2...q_n}$ of the MRSM inequality vanishes for all separable states.}
Specifically, any separable state can be written as a convex combination of product pure states. And for a separable pure state, one can always find a $ \tau_{q_1(q_2\cdots q_{i-1}q_{i+1}...q_{n})}$  equal to $\tau_{q_1(q_2\cdots q_n)}$, in which  case $\tau_{q_1q_i}=0$. For example, without loss of generality, consider the   separable state $|\psi\rangle_{n-1}\otimes | 0\rangle$, with $|\psi\rangle_{n-1}$ being an $n-1$-qubit state. Then from Eq.~(\ref{Eq.b11}), we have
\begin{align}
\tau_{q_1(q_2\cdots q_{n-1})} \! &=  \! \sum_{j=2}^{n-1}\tau_{q_1q_j} \! +\!  \max \limits_{\vec{j}^m:S_{m+1}}\left\{ \sum_{m=3}^{n-2}\!  \tau_{q_1q_{j^m_1}... q_{j^m_{m-1}}} \right\}+ \tau_{q_1q_2\cdots q_{n-1}} \nonumber \\
&=  \sum_{j=2}^n\tau_{q_1q_j} \! +\! \max \limits_{\vec{j}^m:S_{m+1}}\left\{ \sum_{m=3}^{n-1}\!  \tau_{q_1q_{j^m_1}... q_{j^m_{m-1}}}\right \} \nonumber  \\
&= \tau_{q_1(q_2\cdots q_n)}.
\label{Eq.bb16}
\end{align}
}

\medskip

 {
 \section{Comparison of  SM inequalities}\label{sec:compare}
Here, we compare the tightness of the WSM, MRSM and original SM inequalities for four-qubit pure states.
Specifically, for  four-qubit states, the MRSM inequality reduces to
   \begin{align}
\tau_{A(BCD)} \! \geq \! \tau_{AB} \!+\! \tau_{AC} \!+\! \tau_{AD} \!+\! \max  \{ {\cal \tau}_{ABC}\!+\!{\cal \tau}_{ABD}\!+\!{\cal \tau}_{ACD}\},
\label{Eq.b18}
\end{align}
which is  obviously tighter than the WSM inequality, i.e., Eq.~(\ref{Eq.b9}). On the other hand, for four-qubit states, the SM inequality in Ref. \cite{PhysRevLett.113.110501} reduces to
   \begin{align}
\tau_{A(BCD)} \! \geq \! \tau_{AB} \!+\! \tau_{AC} \!+\! \tau_{AD} \!+\!   {\cal \tau}_{ABC}^{3/2} \!+\!{\cal \tau}_{ABD}^{3/2}\!+\!{\cal \tau}_{ACD}^{3/2}.
\label{Eq.b19}
\end{align}
To compare the MRSM and original SM inequalities,  it is sufficient to compare $T_1\equiv \max  \{ {\cal \tau}_{ABC}\!+\!{\cal \tau}_{ABD}\!+\!{\cal \tau}_{ACD}\}$ and $T_2\equiv {\cal \tau}_{ABC}^{3/2} \!+\!{\cal \tau}_{ABD}^{3/2}\!+\!{\cal \tau}_{ACD}^{3/2}$.
}

Recall that Verstraete \emph{et al.}  proposed an insightful classification for four-qubit states into nine groups \cite{PhysRevA.65.052112}.
Specifically,  any four-qubit pure state $| \psi \rangle$ can be obtained as
 \begin{align}
   |\psi\rangle=(A_1\otimes A_2\otimes A_3\otimes A_4)|G^x\rangle,
    \label{Eq.25}
  \end{align}
  Here, $\{A_k\} $ represents a set of Stochastic Local Operations and Classical Communication (SLOCC) operations within $ \mathrm{SL}(2,\mathbb{C})$, where each operation satisfies $\det(A_k)=1$.
  The (unnormalized) states $|G^x\rangle$, indexed by $x=1,...9$, represent nine normal-form families of states,  which are given by
\[
\begin{aligned}
|G_{abcd}^{1}\rangle &= \mbox{$\frac{a + d}{2}(|0000\rangle + |1111\rangle)
+ \frac{a - d}{2}(|0011\rangle + |1100\rangle)$} \\
&+ \mbox{$\frac{b + c}{2}(|0101\rangle + |1010\rangle)
+ \frac{b - c}{2}(|0110\rangle + |1001\rangle)$}, \\
|G_{abc}^{2}\rangle &= \mbox{$\frac{a + b}{2}(|0000\rangle + |1111\rangle)
+ \frac{a - b}{2}(|0011\rangle + |1100\rangle)$} \\
&+ \mbox{$c(|0101\rangle + |1010\rangle) + |0110\rangle$}, \\
|G_{ab}^{3}\rangle &= a(|0000\rangle + |1111\rangle) + b(|0101\rangle + |1010\rangle) \\
&+ |0110\rangle + |0011\rangle, \\
|G_{ab}^{4}\rangle &= a(|0000\rangle + |1111\rangle) \\
&+\mbox{$\frac{a + b}{2}(|0101\rangle + |1010\rangle)
+ \frac{a - b}{2}(|0110\rangle + |1001\rangle)$} \\
&+ \mbox{$\frac{i}{\sqrt{2}}(|0001\rangle + |0010\rangle
+ |0111\rangle + |1011\rangle)$},  \\
|G_{a}^{5}\rangle &= a(|0000\rangle + |0101\rangle + |1010\rangle + |1111\rangle)\\
&+ i|0001\rangle + |0110\rangle - i|1011\rangle, \\
|G_{a}^{6}\rangle &= a(|0000\rangle + |1111\rangle) + |0011\rangle
+ |0101\rangle + |0110\rangle, \\
|G^{7}\rangle &= |0000\rangle + |0101\rangle + |1000\rangle + |1110\rangle, \\
|G^{8}\rangle &= |0000\rangle + |1011\rangle + |1101\rangle + |1110\rangle, \\
|G^{9}\rangle &= |0000\rangle + |0111\rangle,
\end{aligned}
\]
where $a,b,c,d$ are complex parameters with non-negative real parts.
We will use upper bounds of the reduced three-tangles for some typical instances of the normal-form states $|G^x\rangle$, which are given in Refs. \cite{PhysRevLett.113.110501,PhysRevLett.116.049902} and  presented in Table \ref{table1}.

\begin{table}[]
 {
\caption{\label{table1}%
Upper bounds of the reduced three-tangles for the normal-form states $|G^x\rangle$
 }
\begin{ruledtabular}
\begin{tabular}{ccddd}
$|G^x\rangle$ & Upper bounds of the reduced three-tangles\\
\hline
$|G_{abc}^{2}\rangle$ & $\tau_{q_iq_jq_k} \leq \frac{4|c|\sqrt{(a^2 - b^2)(a^{*2} - b^{*2})}}{(|a|^2 + |b|^2 + 2|c|^2 + 1)^2} $\\
$|G_{ab}^{3}\rangle$ & $\tau_{q_1q_2q_3} = \tau_{q_1q_3q_4} = 0, \quad \tau_{q_1q_2q_4}  \leq \frac{4|a||b|}{(1 + |a|^2 + |b|^2)^2}$ \\
$|G_{ab}^{4}\rangle$ & $\tau_{q_iq_jq_k} \leq \frac{2|a^2 - b^2|}{(2 + 3|a|^2 + |b|^2)^2}$  \\
$|G_{a}^{5}\rangle$ & $\tau_{q_1q_2q_3} = \tau_{q_1q_3q_4} \leq \frac{16|a|^2}{(3 + 4|a|^2)^2}, \quad \tau_{q_1q_2q_4} \leq \frac{4}{(3 + 4|a|^2)^2}$ \\
\end{tabular}
\end{ruledtabular}
}
\end{table}

\begin{figure}[]
\centering
\includegraphics[width=8.2cm]{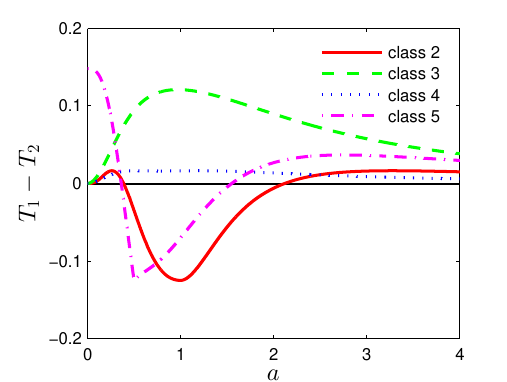}
\caption{(Color online)   {Dependence of $T_1-T_2$ on the  parameter $a$ for the states: $|G_{abc}^{2}\rangle$ with $c=b+1=a$ and $b \ge 0$ (red solid line), $|G_{ab}^{3}\rangle$ with $b=a/4$ (green dashed line), $|G_{ab}^{4}\rangle$ with $b=a/2$ (blue dotted line), $|G_{a}^{5}\rangle$ (magenta dot-dashed line).}}
\label{f1}
\end{figure}

Fig.~\ref{f1} illustrates  how the tightness indicator $T_1-T_2$ depends on  the parameter $a$; the  other parameters are chosen as in  Ref.~\cite{PhysRevLett.113.110501}.
Note that $T_1-T_2 >0$ signifies that the MRSM is tighter than the original SM inequality, and vice versa when $T_1-T_2 <0$.
Fig.~\ref{f1} reveals that for classes 3 and 4, the MRSM inequality is tighter across the entire range of parameters. In contrast, for classes 2 and 5, the MRSM inequality is tighter at the extremes of the parameter
$a$, but becomes looser than the original SM inequality in the intermediate range.

 \section{Example of four-qubit mixed states}\label{sec:fourmix}
 {
Since the residual entanglement of the MRSM state is defined via the convex roof construction,   {whose exact evaluation is generally difficult, it becomes meaningful to derive suitable lower and upper bounds \cite{PhysRevLett.95.040504,PhysRevLett.98.140505,PhysRevA.78.042308}. Equation (\ref{Eq.b13}) shows that directly subtracting the convex roofs of the one-tangle and the
$m$-partite residual entanglement provides an upper bound. On the other hand, analytical solutions can be obtained for low-rank mixed states  \cite{PhysRevA.72.022309,PhysRevLett.97.260502,PhysRevA.77.032310}.
Building on this progress,}
 we  investigate  the MRSM inequality for squared concurrence by considering the following four-qubit mixed state:
\begin{align}
{\rho_{p}} = (1 - p)\left| {\rm GHZ_4} \right\rangle \left\langle {\rm GHZ_4} \right| + p\left| {\rm GHZ_3} \right\rangle \left\langle {\rm GHZ_3} \right| \otimes  {\left| -\right\rangle \left\langle - \right|},
\label{Eq.b25}
\end{align}
where $0\le p \le 1$, $\left| {\rm GHZ_4} \right\rangle  = 1/\sqrt 2 \left( {\left| {0000} \right\rangle  + \left| {1111} \right\rangle } \right)$, $\left| {\rm GHZ_3} \right\rangle $ is the three-qubit Greenberger-Horne-Zeilinger (GHZ) state $1/\sqrt 2 \left( {\left| {000} \right\rangle  + \left| {111} \right\rangle } \right)$ and  {$\left| -\right\rangle=1/\sqrt 2 \left( {\left| {0} \right\rangle  - \left| {1} \right\rangle } \right)$.}
It is obvious that the tangles  $\tau_{A(BCD)}$ of the states $\left| {\rm GHZ_4} \right\rangle $ and $\left| {\rm GHZ_3} \right\rangle\left| {-} \right\rangle  $ are  both equal to  1.
 The  two-qubit tangles $\tau_{AB}$, $\tau_{AC}$ and $\tau_{AD}$ for these two states all vanish. The key difference  is that $\left| {\rm GHZ_4} \right\rangle $ has no three-tangle in any of its reduced subsystems, thus exhibiting a four-partite residual entanglement of 1. In contrast, for $\left| {\rm GHZ_3} \right\rangle\left| {-} \right\rangle  $, the maximum three-tangle among its reduced subsystems is $\tau_{ABC}=1$,  meaning it possesses no four-partite residual entanglement.
 In other words, this mixed state eliminates the contribution of the two-qubit tangle, thereby highlighting the role of tripartite residual entanglement within the MRSM inequality.

 {
For this four-qubit mixed state,  a pure state can be constructed from its eigenvectors as
\begin{align}
{\left| \psi \right\rangle_{p}} = \sqrt{1 - p}\left| {\rm GHZ_4} \right\rangle  +\sqrt{ p}e^{i\phi}\left| {\rm GHZ_3} \right\rangle  \left| -\right\rangle ,
\label{Eq.c21}
\end{align}
such that any pure-state decomposition can be expressed in this form.
From the reduced density matrix, it follows that  the  two-qubit tangles $\tau_{AB}$, $\tau_{AC}$ and $\tau_{AD}$  all vanish, and the tripartite entanglement is exclusively confined to the subsystem $\rho_{ABC}$.
Consequently,
  the residual entanglement of the MRSM inequality reduce to
\begin{align}
\tau _{ABCD}(\left| \psi \right\rangle_{p}) = \tau _{A\left( {BCD} \right)} - \tau _{ABC}.
\label{Eq.b27}
\end{align}
}

In addition, the reduced subsystem $\rho_{ABC}$ of $\left| \psi \right\rangle_{p}$ is a  GHZ-symmetric state \cite{PhysRevLett.108.020502,PhysRevA.89.022312}, for which the three-tangle can be analytically computed \cite{PhysRevLett.108.230502}.
Specifically, any such state $\rho^S$ can be parameterized by coordinates $(x,y)$, defined as follows:
\begin{align}
x(\rho^S) &= \frac{1}{2} \left[ \langle {\rm{GHZ}}_+ | \rho^S | {\rm{GHZ}}_+ \rangle - \langle {\rm{GHZ}}_- | \rho^S |{\rm{GHZ}}_- \rangle \right], \nonumber \\
y(\rho^S)& = \frac{1}{\sqrt{3}} \left[ \langle {\rm{GHZ}}_+ | \rho^S | {\rm{GHZ}}_+ \rangle + \langle {\rm{GHZ}}_- | \rho^S | {\rm{GHZ}}_- \rangle - \frac{1}{4} \right] ,
\label{Eq.bb24}
\end{align}
with $|{\rm{GHZ}}_\pm\rangle = (|000\rangle \pm |111\rangle)/\sqrt{2}$.
Then, given a GHZ-symmetric three-qubit state with coordinates $(x_0, y_0)$, the  three-tangle is given by
\begin{align}
\tau_3(x_0, y_0) =
\begin{cases}
0 & \text{for } x_0 < x_0^W \text{ and } y_0 < y_0^W \\
\left(\frac{x_0 - x_0^W}{\frac{1}{2} - x_0^W}\right)^2 = \left(\frac{y_0 - y_0^W}{\frac{\sqrt{3}}{4} - y_0^W}\right)^2 & \text{otherwise}
\end{cases},
\label{Eq.bb25}
\end{align}
where $(x_0^W, y_0^W)$ is the intersection point of the  line joining  the $|{\rm{GHZ}}_+\rangle$ state at $(1/2, \sqrt{3}/4)$ to  $(x_0, y_0)$ with the ``GHZ/W line'', the latter being defined by   the parameterized curve
\begin{align}
x^W = \frac{v^5 + 8v^3}{8(4 - v^2)}, \quad y = \frac{\sqrt{3}}{4} \frac{4 - v^2 - v^4}{4 - v^2},
\label{Eq.bb26}
\end{align}
where $y \geq \frac{1}{2\sqrt{3}} \text{ and } -1 \leq v \leq 1$.

Using Eq.~(\ref{Eq.bb24}), the coordinates of $\rho_{ABC}$ are determined to be $(p/2,\sqrt{3}/4)$.
Then direct calculation of Eq.~(\ref{Eq.bb25}) reveals that  the three-tangle of $\rho_{ABC}$ is $\tau_{ABC}=p^2$.
 {
Determining the four-partite residual entanglement of $\rho_{p}$ reduces to finding the convex roof of the one-tangle.
By applying the method from Ref.~\cite{PhysRevA.77.032310}, we first derive the one-tangle of the superposition state as a function of the relative phase $\phi$:
\begin{equation}
    \tau_{A(BCD)}(p,\phi) = 1 - p(1-p)(1 + \cos 2\phi).
\end{equation}
The tangle of the mixed state is determined by minimizing this quantity over $\phi$, which yields $\tau_{A(BCD)} = 1 - 2p + 2p^2$. Consequently, subtracting the tripartite contribution leads to a four-partite residual entanglement of $(1-p)^2$.
It is also insightful to compare how the four-partite and tripartite entanglements vary with the parameter $p$.
The tripartite entanglement of $\rho_p$ is confined to the $ABC$ subsystem, which itself is also a GHZ-symmetric state. Quantified by the three-tangle, this entanglement remains $p^2$.
Fig.~\ref{f2} plots how the four-partite residual entanglement $\tau _{ABCD}$ and the maximum reduced  three-tangle $\tau _{ABC}$ depend on the parameter $p$.
 As shown in the figure,  $\tau _{ABCD}$  decrease as $\tau _{ABC}$ increase, confirming the validity of the MRSM inequality.
}

\begin{figure}[]
\centering
\includegraphics[width=8.2cm]{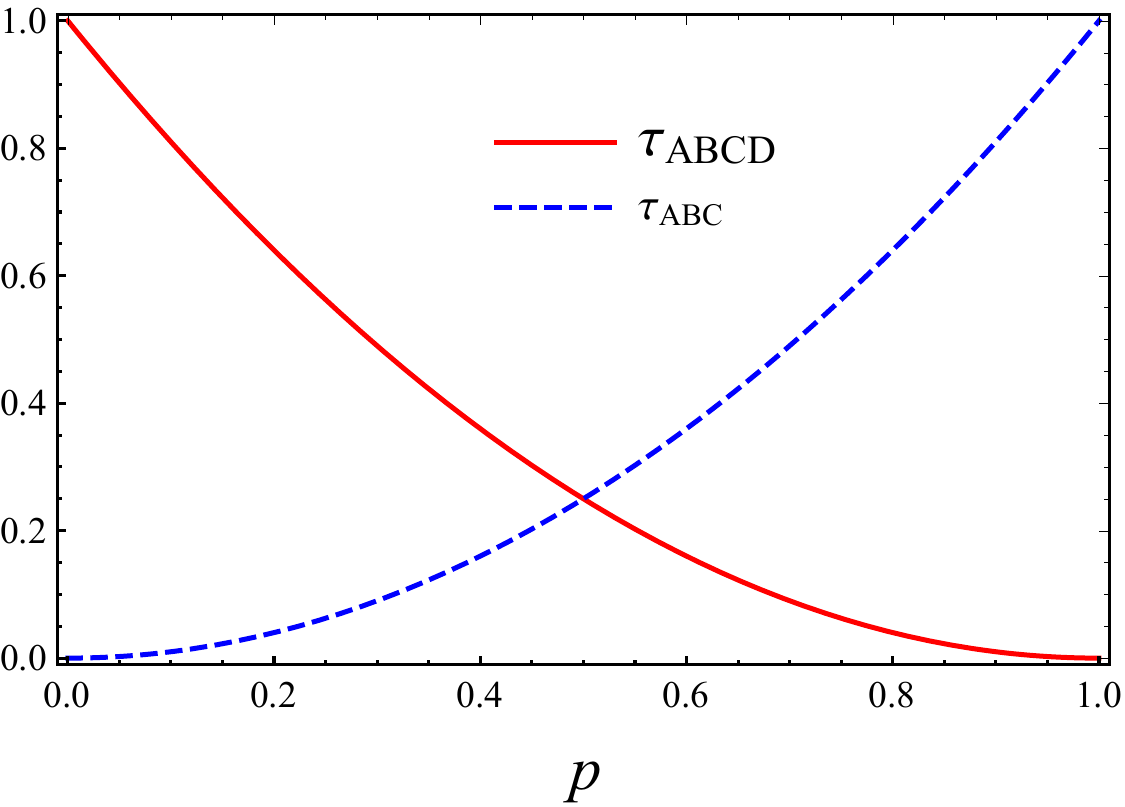}
\caption{(Color online)    {Dependence of the four-partite residual entanglement $\tau _{ABCD}$ (red solid line) and the maximum reduced  three-tangle $\tau _{ABC}$ (blue dashed line) on the  parameter $p$  for the mixed state ${\rho_{p}}$ [Eq.~\eqref{Eq.b25}].}}
\label{f2}
\end{figure}

 {
 \section{Example of  five-qubit pure states}\label{sec:fivepure}
The MRSM inequality is defined for an $n$-qubit system; however, the right-hand side requires recursively invoking the
$m$-partite residual entanglement, which is defined via the convex roof construction. Consequently, the computational complexity increases dramatically with the number of qubits. Here, we provide an example of a five-qubit pure state, which is constructed as follows:
\begin{align}
\left | \Psi  \right \rangle _{p} =\sqrt{1-p} \left|{\rm GHZ}_4\right\rangle \left | 0 \right \rangle +\sqrt{p}\left|{\rm GHZ}_3\right\rangle\left | - \right \rangle \left | 1 \right \rangle.
\label{Eq.c26}
\end{align}
For this state, the reduced density matrices reveal that all reduced two-qubit tangles involving the first qubit vanish. The tripartite entanglement is confined to the subsystem $\rho_{123}$. Furthermore, it can also be seen that the one-tangle (of the first qubit) for the reduced systems $\rho_{1245}$ and  $\rho_{1345}$ is zero, implying the absence of four-partite entanglement.
Consequently, the residual entanglement for this state reduces to
\begin{align}
\tau _{12345}(\left| \Psi \right\rangle_{p}) =\tau _{1\left ( 2345 \right ) } -\max\left \{ \tau _{1234}, \tau _{1235} \right \} -\tau _{123} .
\label{Eq.c27}
\end{align}
Here, $\tau _{1\left ( 2345 \right ) }=1$,  $\tau _{1 234 }=(1-p)^2$ and $\tau _{1 23 }=p^2$, since the subsystem $\rho_{1234}$ corresponds exactly to the four-qubit mixed state example $\rho_p$ defined in Eq.~(\ref{Eq.b25}). Consequently, determining the five-partite residual entanglement reduces to finding the convex roof of
 $\tau _{1235}$. By following a procedure similar to that used for
  $\tau _{1234}$, which involves the eigendecomposition of $\rho _{1235}$, we obtain the result $2p(1-p)$.
 }

 \begin{figure}[]
\centering
\includegraphics[width=8.2cm]{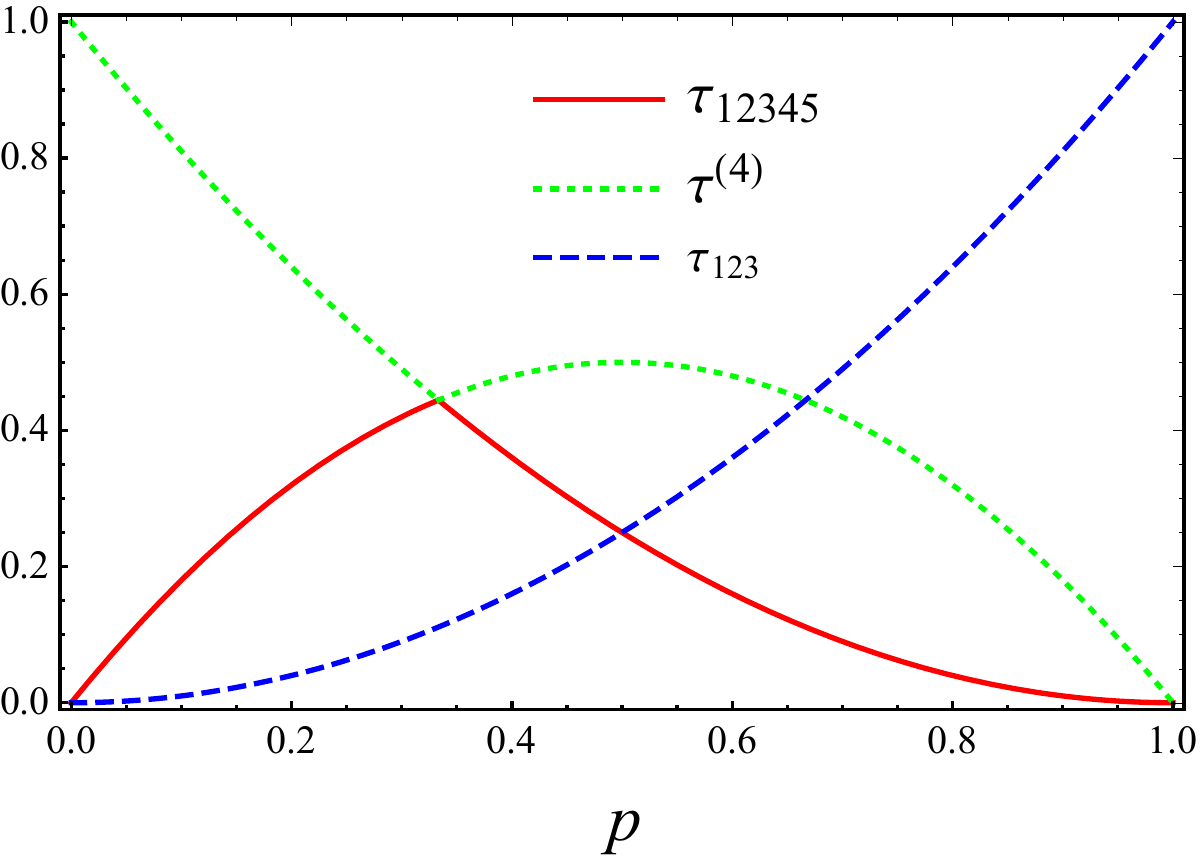}
\caption{(Color online)    {Dependence of the five-partite entanglement $\tau_{12345}$ (red solid line), the maximum  four-partite entanglement $\tau^{(4)}$ (green dotted line) and the tripartite entanglement $\tau_{123}$ (blue dashed line) on the  parameter $p$  for the  state $\left | \Psi  \right \rangle _{p}$ [Eq.~\eqref{Eq.c26}].}}
\label{f3}
\end{figure}

 {
Fig.~\ref{f3} plots  the dependence of the five-partite entanglement $\tau_{12345}$, the maximum  four-partite entanglement $\tau^{(4)}\equiv \max\left \{ \tau _{1234}, \tau _{1235} \right \}$ and the tripartite entanglement $\tau_{123}$ on the  parameter $p$  for the state $\left | \Psi  \right \rangle _{p} $.
It can be seen that for the limiting case $p=0$, the state becomes $\left | \Psi  \right \rangle _{0}=\left|{\rm GHZ}_4\right\rangle\left | 0 \right \rangle$, which possesses only four-partite entanglement. Conversely, when
  $p=1$, the state becomes $\left | \Psi  \right \rangle _{1}=\left|{\rm GHZ}_3\right\rangle\left | - \right \rangle \left | 1 \right \rangle$, which possesses only tripartite entanglement. As $p$
 varies from 0 to 1, the tripartite entanglement increases monotonically, which is consistent with intuition.
    However, the maximum four-partite entanglement does not decrease monotonically; this is because, although
     $\tau_{1234}$ decreases,  $\tau_{1235}$ may increase. Notably,  five-partite entanglement $\tau_{12345}$ emerges in any non-trivial superposition  of
      $\left | \Psi  \right \rangle _{0}$ and $\left | \Psi  \right \rangle _{1}$, even though neither of the individual component states possesses five-partite entanglement.
}

 \section{Discussion and conclusion}\label{sec:concl}
This paper addresses a long-standing question \cite{PhysRevLett.113.110501}: can genuine $m$-partite entanglement  be incorporated into  a generalized CKW-type inequality --- of the form akin to Eq.~(\ref{Eq.3}) --- for multiqubit systems?
Such a {\it strong} monogamy inequality involving $m$-partite entanglement would impose stricter constraints on multiple forms of entanglement distribution compared to the conventional CKW-type monogamy inequality  \cite{PhysRevA.61.052306,PhysRevLett.96.220503,PhysRevLett.117.060501},
which already stimulated significant applications, such as quantum cryptography, many-body spin systems and quantum networks \cite{RevModPhys.81.865,RevModPhys.80.517,PhysRevLett.107.260602,PhysRevA.111.012420,PhysRevA.111.042429}.
A general SM inequality offers a valuable approach for quantifying the essential features of quantum correlations that exclusively arise in scenarios beyond the bipartite case.

Here we first established a universal WSM inequality, by using coefficients rather than exponents to modulate the weight allocated to various $m$-partite contributions, which distinguishes it from the original SM conjecture \cite{PhysRevLett.113.110501}.
We then established the MRSM inequality, which is defined by the maximum $m$-partite entanglement, and leads to a faithful quantifier of residual entanglement.
The derived WSM and MRSM  inequalities  sharpen the existing ones \cite{PhysRevA.61.052306,PhysRevLett.96.220503}, and are  proven analytically for arbitrary multiqubit states.  We showed  that the WSM and MRSM inequalities, initially formulated for squared concurrence, can be naturally generalized to other entanglement measures.
Furthermore, we compared the tightness of different SM inequalities. Finally, we provided {examples using a four-qubit mixed state family and a five-qubit pure state family to demonstrate the validity of the MRSM inequality. Through these examples, we characterized the trade-off relations among entanglement components involving different numbers of qubits.}

 Our results represent a significant step forward in the quantitative quest to understand the distribution of entanglement in complex quantum systems. In future studies, sharper monogamy inequalities than the WSM and MRSM  inequalities might be developed by assigning more weight to  $m$-partite contributions. Furthermore, it would be interesting to analyze the operational significance of the $m$-partite entanglement emerging from the WSM and MRSM  inequalities, and on the other hand to investigate the validity of new monogamy inequalities in higher dimensional systems.
 \vskip0.1cm

\begin{acknowledgements}
 This work was supported by the National Science
Foundation of China under (Grants No. 12475009 and
No. 12075001), Anhui Provincial Key Research and
Development Plan (Grant No. 2022b13020004), Anhui
Province Science and Technology Innovation Project
(Grant No. 202423r06050004),  Anhui Provincial
University Scientific Research Major Project (Grant
No. 2024AH040008), Anhui Provincial Department of Industry and Information Technology (Grant No. JB24044), Anhui Province Natural Science Foundation (Grant  No. 202508140141) and the UK
Research and Innovation  (EPSRC Grant No. EP/X010929/1).
\end{acknowledgements}

\medskip

\appendix
\hypertarget{app1}{}
\section*{Appendix: Proof of Theorem 1}
\begin{proof}
 The difference between left- and right-hand sides of Eq.~(\ref{Eq.8}) for any $n$-qubit pure state  is given by
\begin{align}
\tau_{q_1q_2...q_n} :=& \tau_{q_1(q_2\cdots q_n)} \! - \! \sum_{j=2}^n\tau_{q_1q_j}- \sum_{m=3}^{n-1}\binom{n-1}{m-1} ^{-1} \! \sum_{\vec{j}^m:S_n} \tau_{q_1q_{j^m_1}... q_{j^m_{m-1}}}.
\label{Eq.9}
\end{align}
And  for an arbitrary mixed state $\rho$ it is expressed as
\begin{align}
\tau_{q_1q_2...q_n}:=\inf_{\{p_i, |\psi_i\rangle\}} \sum_i p_i \mathcal{\tau}_{q_1q_2...q_n}(|\psi_i \rangle),
\label{Eq.10}
\end{align}
where the infimum runs over all pure-state decompositions of $\rho=\sum_i p_i |\psi_i\rangle \langle \psi_i|$. Then the $m$-partite  $\tau_{q_1q_{j^m_1}... q_{j^m_{m-1}}}$ is generated via Eqs.~(\ref{Eq.9}) and (\ref{Eq.10})   for corresponding reduced state $\rho_{q_1q_{j^m_1}... q_{j^m_{m-1}}}$ in a recursive manner.
In the following we will demonstrate the validity of Theorem 1 by employing the method of mathematical induction.

For any four-qubit pure state $|\psi \rangle_{ABCD}$, from Eq.~(\ref{Eq.7}), we have
 \begin{align}
\tau_{A(BCD)} &\geq \tau_{AB} + \tau_{A(CD)}     \nonumber         \\
&\geq \tau_{AB} + \tau_{AC}+ \tau_{AD}+\tau_{ACD}.
\label{Eq.11}
\end{align}
The second inequality follows from the fact that, for mixed states,
 \begin{align}
\tau_{ACD} \le \tau_{A(CD)}-\tau_{AC}- \tau_{AD},
\label{Eq.12}
\end{align}
 where $\tau_{ACD}$ denotes the three-tangle,  defined for mixed states through  the  convex roof construction of  Eq.~(\ref{Eq.2}).
To further understand, consider the pure states $|\psi_i\rangle$ belonging to an optimal decomposition of $\rho_{ACD}$, that is, a decomposition that minimizes $\langle \tau_{A(CD)} \rangle= \sum_i p_i \tau_{A(CD)}(|\psi_i\rangle)$. Then $\tau_{ACD} \le \langle \tau_{A(CD)}- \tau_{AC}- \tau_{AD} \rangle=\langle \tau_{A(CD)} \rangle- \langle\tau_{AC} \rangle- \langle \tau_{AD} \rangle \le   \tau_{A(CD)}  - \tau_{AC}- \tau_{AD}$, since both $\tau_{AC}$ and $\tau_{AD}$ are   convex functions on  the set of density matrices \cite{PhysRevLett.80.2245}.
Similar to Eq.~(\ref{Eq.11}), it follows that
 \begin{align}
\tau_{A(BCD)} &\geq \tau_{AC} + \tau_{A(BD)}   \geq \tau_{AC} + \tau_{AB}+ \tau_{AD}+\tau_{ABD}.
\label{Eq.13}
\end{align}
 \begin{align}
\tau_{A(BCD)} \geq \tau_{AB} + \tau_{AC}+ \tau_{AD}+\tau_{ABC}.
\label{Eq.14}
\end{align}
Combing Eqs.~(\ref{Eq.11}), (\ref{Eq.13}) and (\ref{Eq.14}), we obtain
 \begin{align}
\tau_{A(BCD)} \! \geq \! \tau_{AB} \!+\! \tau_{AC} \!+\! \tau_{AD} \!+\! \frac{1}{3} ( \tau_{ABC}+\tau_{ABD}+\tau_{ACD}).
\label{Eq.15}
\end{align}
Without loss of generality, assuming  for arbitrary ($n-1$)-qubit states,   the following WSM inequality holds
\begin{align}
\tau_{q_1(q_2\cdots q_{n-1})} \! \geq \! \sum_{j=2}^{n-1}\tau_{q_1q_j} \! +\! \sum_{m=3}^{n-2} \! \binom{n-2}{m-1} ^{-1} \! \!\!  \sum_{\vec{j}^m:S_n\setminus n} \! \!  \tau_{q_1q_{j^m_1}... q_{j^m_{m-1}}},
\label{Eq.16}
\end{align}
where $S_n\setminus n$ means removing element \emph{n} from $S_n$.
Resorting to Eqs.~(\ref{Eq.9}) and (\ref{Eq.10}), the following inequality is satisfied as
 \begin{align}
\tau_{q_1q_2\cdots q_{n-1}} \le & \tau_{q_1(q_2\cdots q_{n-1})} \! - \! \sum_{j=2}^{n-1}\tau_{q_1q_j} \! \nonumber  \\
&-\! \sum_{m=3}^{n-2} \! \binom{n-2}{m-1} ^{-1} \! \sum_{\vec{j}^m:S_n\setminus n} \tau_{q_1q_{j^m_1}... q_{j^m_{m-1}}}.
\label{Eq.17}
\end{align}
The analysis follows a similar approach to that of Eq.~(\ref{Eq.12}), given that  all $m$-partite residual entanglements are defined via a convex roof construction.
Consequently, we arrive at
\begin{align}
\!\!\tau_{q_1(q_2\cdots q_{n-1})} \! & \geq \! \sum_{j=2}^{n-1}\tau_{q_1q_j} \! +\! \sum_{m=3}^{n-2} \! \binom{n-2}{m-1} ^{-1} \! \!\!\sum_{\vec{j}^m:S_n\setminus n}\!\! \tau_{q_1q_{j^m_1}... q_{j^m_{m-1}}}
\! + \! \tau_{q_1q_2\cdots q_{n-1}} \nonumber \\
 &= \sum_{j=2}^{n-1}\tau_{q_1q_j} \! +\! \sum_{m=3}^{n-1} \! \binom{n-2}{m-1} ^{-1} \! \!\!\sum_{\vec{j}^m:S_n\setminus n} \!\!\tau_{q_1q_{j^m_1}... q_{j^m_{m-1}}}.
\label{Eq.18}
\end{align}
From Eq.~(\ref{Eq.7}), we attain
\begin{align}
\tau_{q_1(q_2\cdots q_{n})} &\geq \tau_{q_1q_i} + \tau_{q_1(q_2...q_{i-1}q_{i+1}...q_n)} \nonumber \\
&\ge \tau_{q_1q_i} + \sum_{\substack{j=2 \\ j \neq i}}^{n}\tau_{q_1q_j} \!   +\! \sum_{m=3}^{n-1} \! \binom{n-2}{m-1} ^{-1} \! \sum_{\vec{j}^m : S_n\setminus i} \tau_{q_1q_{j^m_1}... q_{j^m_{m-1}}} \nonumber \\
&=\sum_{j=2}^{n}\tau_{q_1q_j}+\! \sum_{m=3}^{n-1} \! \binom{n-2}{m-1} ^{-1} \! \sum_{\vec{j}^m: S_n\setminus i} \tau_{q_1q_{j^m_1}... q_{j^m_{m-1}}}.
\label{Eq.19}
\end{align}
For distinct $i \in S_n$, we obtain $n-1$ inequalities with identical left-hand sides. Averaging the right-hand sides of these inequalities, it yields
\begin{align}
\tau_{q_1(q_2\cdots q_{n})} & \geq
\sum_{j=2}^{n}\tau_{q_1q_j} \nonumber  \\
&+\! \sum_{m=3}^{n-1} \! \binom{n-2}{m-1} ^{-1} \! \frac{1 }{n-1}  \!\sum_{i=2}^n \sum_{\vec{j}^m: S_n\setminus i} \! \tau_{q_1q_{j^m_1}... q_{j^m_{m-1}}}.
\label{Eq.20}
\end{align}
Note that the indices of the  terms in the sum   $\sum_{i=2}^n \sum_{\vec{j}^m: S_n\setminus i} \tau_{q_1q_{j^m_1}... q_{j^m_{m-1}}}$ form  a list $L_1$, which can be constructed as follows: firstly, select $n-2$ elements from  $S_n$, and then choose $m-1$ elements from those $n-2$ elements; all possible combinations of such choices constitute the List $L_1$. The number of these combinations is given by $(n-1)\binom{n-2}{m-1}$.
On the other hand, the indices of the  terms in the sum $ \sum_{\vec{j}^m: S_n} \tau_{q_1q_{j^m_1}... q_{j^m_{m-1}}}$ form  a set $S_1$, which corresponds to the combinations of  directly selecting $m-1$ elements from $S_n$, and the number of such combinations is expressed as  $\binom{n-1}{m-1}$.
In fact,  the list $L_1$ is exactly the set  $S_1$ after removing repeated elements. And one can show that $(n-1)\binom{n-2}{m-1}=(n-m)\binom{n-1}{m-1}$. That is to say
\begin{align}
\tau_{q_1(q_2\cdots q_{n})} &\geq
\sum_{j=2}^{n}\tau_{q_1q_j}+  \sum_{m=3}^{n-1}  \binom{n-2}{m-1} ^{-1}\frac{n-m }{n-1}   \sum_{\vec{j}^m: S_n} \tau_{q_1q_{j^m_1}... q_{j^m_{m-1}}} \nonumber \\
&=\sum_{j=2}^n\tau_{q_1q_j} + \sum_{m=3}^{n-1}\binom{n-1}{m-1} ^{-1} \! \sum_{\vec{j}^m:S_n} \tau_{q_1q_{j^m_1}... q_{j^m_{m-1}}}.
\label{Eq.21}
\end{align}
This completes the proof.
\end{proof}


\begin{thebibliography}{36}%
\makeatletter
\providecommand \@ifxundefined [1]{%
 \@ifx{#1\undefined}
}%
\providecommand \@ifnum [1]{%
 \ifnum #1\expandafter \@firstoftwo
 \else \expandafter \@secondoftwo
 \fi
}%
\providecommand \@ifx [1]{%
 \ifx #1\expandafter \@firstoftwo
 \else \expandafter \@secondoftwo
 \fi
}%
\providecommand \natexlab [1]{#1}%
\providecommand \enquote  [1]{``#1''}%
\providecommand \bibnamefont  [1]{#1}%
\providecommand \bibfnamefont [1]{#1}%
\providecommand \citenamefont [1]{#1}%
\providecommand \href@noop [0]{\@secondoftwo}%
\providecommand \href [0]{\begingroup \@sanitize@url \@href}%
\providecommand \@href[1]{\@@startlink{#1}\@@href}%
\providecommand \@@href[1]{\endgroup#1\@@endlink}%
\providecommand \@sanitize@url [0]{\catcode `\\12\catcode `\$12\catcode
  `\&12\catcode `\#12\catcode `\^12\catcode `\_12\catcode `\%12\relax}%
\providecommand \@@startlink[1]{}%
\providecommand \@@endlink[0]{}%
\providecommand \url  [0]{\begingroup\@sanitize@url \@url }%
\providecommand \@url [1]{\endgroup\@href {#1}{\urlprefix }}%
\providecommand \urlprefix  [0]{URL }%
\providecommand \Eprint [0]{\href }%
\providecommand \doibase [0]{http://dx.doi.org/}%
\providecommand \selectlanguage [0]{\@gobble}%
\providecommand \bibinfo  [0]{\@secondoftwo}%
\providecommand \bibfield  [0]{\@secondoftwo}%
\providecommand \translation [1]{[#1]}%
\providecommand \BibitemOpen [0]{}%
\providecommand \bibitemStop [0]{}%
\providecommand \bibitemNoStop [0]{.\EOS\space}%
\providecommand \EOS [0]{\spacefactor3000\relax}%
\providecommand \BibitemShut  [1]{\csname bibitem#1\endcsname}%
\let\auto@bib@innerbib\@empty
\bibitem [{\citenamefont {Horodecki}\ \emph {et~al.}(2009)\citenamefont
  {Horodecki}, \citenamefont {Horodecki}, \citenamefont {Horodecki},\ and\
  \citenamefont {Horodecki}}]{RevModPhys.81.865}%
  \BibitemOpen
  \bibfield  {author} {\bibinfo {author} {\bibfnamefont {R.}~\bibnamefont
  {Horodecki}}, \bibinfo {author} {\bibfnamefont {P.}~\bibnamefont
  {Horodecki}}, \bibinfo {author} {\bibfnamefont {M.}~\bibnamefont
  {Horodecki}}, \ and\ \bibinfo {author} {\bibfnamefont {K.}~\bibnamefont
  {Horodecki}},\ }\href {\doibase 10.1103/RevModPhys.81.865} {\bibfield
  {journal} {\bibinfo  {journal} {Rev. Mod. Phys.}\ }\textbf {\bibinfo {volume}
  {81}},\ \bibinfo {pages} {865} (\bibinfo {year} {2009})}\BibitemShut
  {NoStop}%
\bibitem [{\citenamefont {Amico}\ \emph {et~al.}(2008)\citenamefont {Amico},
  \citenamefont {Fazio}, \citenamefont {Osterloh},\ and\ \citenamefont
  {Vedral}}]{RevModPhys.80.517}%
  \BibitemOpen
  \bibfield  {author} {\bibinfo {author} {\bibfnamefont {L.}~\bibnamefont
  {Amico}}, \bibinfo {author} {\bibfnamefont {R.}~\bibnamefont {Fazio}},
  \bibinfo {author} {\bibfnamefont {A.}~\bibnamefont {Osterloh}}, \ and\
  \bibinfo {author} {\bibfnamefont {V.}~\bibnamefont {Vedral}},\ }\href
  {\doibase 10.1103/RevModPhys.80.517} {\bibfield  {journal} {\bibinfo
  {journal} {Rev. Mod. Phys.}\ }\textbf {\bibinfo {volume} {80}},\ \bibinfo
  {pages} {517} (\bibinfo {year} {2008})}\BibitemShut {NoStop}%
\bibitem [{\citenamefont {Bennett}\ \emph {et~al.}(1996)\citenamefont
  {Bennett}, \citenamefont {DiVincenzo}, \citenamefont {Smolin},\ and\
  \citenamefont {Wootters}}]{PhysRevA.54.3824}%
  \BibitemOpen
  \bibfield  {author} {\bibinfo {author} {\bibfnamefont {C.~H.}\ \bibnamefont
  {Bennett}}, \bibinfo {author} {\bibfnamefont {D.~P.}\ \bibnamefont
  {DiVincenzo}}, \bibinfo {author} {\bibfnamefont {J.~A.}\ \bibnamefont
  {Smolin}}, \ and\ \bibinfo {author} {\bibfnamefont {W.~K.}\ \bibnamefont
  {Wootters}},\ }\href {\doibase 10.1103/PhysRevA.54.3824} {\bibfield
  {journal} {\bibinfo  {journal} {Phys. Rev. A}\ }\textbf {\bibinfo {volume}
  {54}},\ \bibinfo {pages} {3824} (\bibinfo {year} {1996})}\BibitemShut
  {NoStop}%
\bibitem [{\citenamefont {Hill}\ and\ \citenamefont
  {Wootters}(1997)}]{PhysRevLett.78.5022}%
  \BibitemOpen
  \bibfield  {author} {\bibinfo {author} {\bibfnamefont {S.~A.}\ \bibnamefont
  {Hill}}\ and\ \bibinfo {author} {\bibfnamefont {W.~K.}\ \bibnamefont
  {Wootters}},\ }\href {\doibase 10.1103/PhysRevLett.78.5022} {\bibfield
  {journal} {\bibinfo  {journal} {Phys. Rev. Lett.}\ }\textbf {\bibinfo
  {volume} {78}},\ \bibinfo {pages} {5022} (\bibinfo {year}
  {1997})}\BibitemShut {NoStop}%
\bibitem [{\citenamefont {Wootters}(1998)}]{PhysRevLett.80.2245}%
  \BibitemOpen
  \bibfield  {author} {\bibinfo {author} {\bibfnamefont {W.~K.}\ \bibnamefont
  {Wootters}},\ }\href {\doibase 10.1103/PhysRevLett.80.2245} {\bibfield
  {journal} {\bibinfo  {journal} {Phys. Rev. Lett.}\ }\textbf {\bibinfo
  {volume} {80}},\ \bibinfo {pages} {2245} (\bibinfo {year}
  {1998})}\BibitemShut {NoStop}%
\bibitem [{\citenamefont {Rungta}\ \emph {et~al.}(2001)\citenamefont {Rungta},
  \citenamefont {Bu\ifmmode~\check{z}\else \v{z}\fi{}ek}, \citenamefont
  {Caves}, \citenamefont {Hillery},\ and\ \citenamefont
  {Milburn}}]{PhysRevA.64.042315}%
  \BibitemOpen
  \bibfield  {author} {\bibinfo {author} {\bibfnamefont {P.}~\bibnamefont
  {Rungta}}, \bibinfo {author} {\bibfnamefont {V.}~\bibnamefont
  {Bu\ifmmode~\check{z}\else \v{z}\fi{}ek}}, \bibinfo {author} {\bibfnamefont
  {C.~M.}\ \bibnamefont {Caves}}, \bibinfo {author} {\bibfnamefont
  {M.}~\bibnamefont {Hillery}}, \ and\ \bibinfo {author} {\bibfnamefont
  {G.~J.}\ \bibnamefont {Milburn}},\ }\href {\doibase
  10.1103/PhysRevA.64.042315} {\bibfield  {journal} {\bibinfo  {journal} {Phys.
  Rev. A}\ }\textbf {\bibinfo {volume} {64}},\ \bibinfo {pages} {042315}
  (\bibinfo {year} {2001})}\BibitemShut {NoStop}%
\bibitem [{\citenamefont {Vidal}\ and\ \citenamefont
  {Werner}(2002)}]{PhysRevA.65.032314}%
  \BibitemOpen
  \bibfield  {author} {\bibinfo {author} {\bibfnamefont {G.}~\bibnamefont
  {Vidal}}\ and\ \bibinfo {author} {\bibfnamefont {R.~F.}\ \bibnamefont
  {Werner}},\ }\href {\doibase 10.1103/PhysRevA.65.032314} {\bibfield
  {journal} {\bibinfo  {journal} {Phys. Rev. A}\ }\textbf {\bibinfo {volume}
  {65}},\ \bibinfo {pages} {032314} (\bibinfo {year} {2002})}\BibitemShut
  {NoStop}%
\bibitem [{\citenamefont {Dong}\ \emph {et~al.}(2022)\citenamefont {Dong},
  \citenamefont {Wei}, \citenamefont {Song}, \citenamefont {Wang},\ and\
  \citenamefont {Ye}}]{PhysRevA.106.042415}%
  \BibitemOpen
  \bibfield  {author} {\bibinfo {author} {\bibfnamefont {D.-D.}\ \bibnamefont
  {Dong}}, \bibinfo {author} {\bibfnamefont {G.-B.}\ \bibnamefont {Wei}},
  \bibinfo {author} {\bibfnamefont {X.-K.}\ \bibnamefont {Song}}, \bibinfo
  {author} {\bibfnamefont {D.}~\bibnamefont {Wang}}, \ and\ \bibinfo {author}
  {\bibfnamefont {L.}~\bibnamefont {Ye}},\ }\href {\doibase
  10.1103/PhysRevA.106.042415} {\bibfield  {journal} {\bibinfo  {journal}
  {Phys. Rev. A}\ }\textbf {\bibinfo {volume} {106}},\ \bibinfo {pages}
  {042415} (\bibinfo {year} {2022})}\BibitemShut {NoStop}%
\bibitem [{\citenamefont {Dong}\ \emph {et~al.}(2023)\citenamefont {Dong},
  \citenamefont {Song}, \citenamefont {Fan}, \citenamefont {Ye},\ and\
  \citenamefont {Wang}}]{PhysRevA.107.052403}%
  \BibitemOpen
  \bibfield  {author} {\bibinfo {author} {\bibfnamefont {D.-D.}\ \bibnamefont
  {Dong}}, \bibinfo {author} {\bibfnamefont {X.-K.}\ \bibnamefont {Song}},
  \bibinfo {author} {\bibfnamefont {X.-G.}\ \bibnamefont {Fan}}, \bibinfo
  {author} {\bibfnamefont {L.}~\bibnamefont {Ye}}, \ and\ \bibinfo {author}
  {\bibfnamefont {D.}~\bibnamefont {Wang}},\ }\href {\doibase
  10.1103/PhysRevA.107.052403} {\bibfield  {journal} {\bibinfo  {journal}
  {Phys. Rev. A}\ }\textbf {\bibinfo {volume} {107}},\ \bibinfo {pages}
  {052403} (\bibinfo {year} {2023})}\BibitemShut {NoStop}%
\bibitem [{\citenamefont {Coffman}\ \emph {et~al.}(2000)\citenamefont
  {Coffman}, \citenamefont {Kundu},\ and\ \citenamefont
  {Wootters}}]{PhysRevA.61.052306}%
  \BibitemOpen
  \bibfield  {author} {\bibinfo {author} {\bibfnamefont {V.}~\bibnamefont
  {Coffman}}, \bibinfo {author} {\bibfnamefont {J.}~\bibnamefont {Kundu}}, \
  and\ \bibinfo {author} {\bibfnamefont {W.~K.}\ \bibnamefont {Wootters}},\
  }\href {\doibase 10.1103/PhysRevA.61.052306} {\bibfield  {journal} {\bibinfo
  {journal} {Phys. Rev. A}\ }\textbf {\bibinfo {volume} {61}},\ \bibinfo
  {pages} {052306} (\bibinfo {year} {2000})}\BibitemShut {NoStop}%
\bibitem [{\citenamefont {Osborne}\ and\ \citenamefont
  {Verstraete}(2006)}]{PhysRevLett.96.220503}%
  \BibitemOpen
  \bibfield  {author} {\bibinfo {author} {\bibfnamefont {T.~J.}\ \bibnamefont
  {Osborne}}\ and\ \bibinfo {author} {\bibfnamefont {F.}~\bibnamefont
  {Verstraete}},\ }\href {\doibase 10.1103/PhysRevLett.96.220503} {\bibfield
  {journal} {\bibinfo  {journal} {Phys. Rev. Lett.}\ }\textbf {\bibinfo
  {volume} {96}},\ \bibinfo {pages} {220503} (\bibinfo {year}
  {2006})}\BibitemShut {NoStop}%
\bibitem [{\citenamefont {Ou}\ and\ \citenamefont
  {Fan}(2007)}]{PhysRevA.75.062308}%
  \BibitemOpen
  \bibfield  {author} {\bibinfo {author} {\bibfnamefont {Y.-C.}\ \bibnamefont
  {Ou}}\ and\ \bibinfo {author} {\bibfnamefont {H.}~\bibnamefont {Fan}},\
  }\href {\doibase 10.1103/PhysRevA.75.062308} {\bibfield  {journal} {\bibinfo
  {journal} {Phys. Rev. A}\ }\textbf {\bibinfo {volume} {75}},\ \bibinfo
  {pages} {062308} (\bibinfo {year} {2007})}\BibitemShut {NoStop}%
\bibitem [{\citenamefont {de~Oliveira}\ \emph {et~al.}(2014)\citenamefont
  {de~Oliveira}, \citenamefont {Cornelio},\ and\ \citenamefont
  {Fanchini}}]{PhysRevA.89.034303}%
  \BibitemOpen
  \bibfield  {author} {\bibinfo {author} {\bibfnamefont {T.~R.}\ \bibnamefont
  {de~Oliveira}}, \bibinfo {author} {\bibfnamefont {M.~F.}\ \bibnamefont
  {Cornelio}}, \ and\ \bibinfo {author} {\bibfnamefont {F.~F.}\ \bibnamefont
  {Fanchini}},\ }\href {\doibase 10.1103/PhysRevA.89.034303} {\bibfield
  {journal} {\bibinfo  {journal} {Phys. Rev. A}\ }\textbf {\bibinfo {volume}
  {89}},\ \bibinfo {pages} {034303} (\bibinfo {year} {2014})}\BibitemShut
  {NoStop}%
\bibitem [{\citenamefont {Bai}\ \emph {et~al.}(2014)\citenamefont {Bai},
  \citenamefont {Xu},\ and\ \citenamefont {Wang}}]{PhysRevLett.113.100503}%
  \BibitemOpen
  \bibfield  {author} {\bibinfo {author} {\bibfnamefont {Y.-K.}\ \bibnamefont
  {Bai}}, \bibinfo {author} {\bibfnamefont {Y.-F.}\ \bibnamefont {Xu}}, \ and\
  \bibinfo {author} {\bibfnamefont {Z.~D.}\ \bibnamefont {Wang}},\ }\href
  {\doibase 10.1103/PhysRevLett.113.100503} {\bibfield  {journal} {\bibinfo
  {journal} {Phys. Rev. Lett.}\ }\textbf {\bibinfo {volume} {113}},\ \bibinfo
  {pages} {100503} (\bibinfo {year} {2014})}\BibitemShut {NoStop}%
\bibitem [{\citenamefont {Zhu}\ and\ \citenamefont
  {Fei}(2014)}]{PhysRevA.90.024304}%
  \BibitemOpen
  \bibfield  {author} {\bibinfo {author} {\bibfnamefont {X.-N.}\ \bibnamefont
  {Zhu}}\ and\ \bibinfo {author} {\bibfnamefont {S.-M.}\ \bibnamefont {Fei}},\
  }\href {\doibase 10.1103/PhysRevA.90.024304} {\bibfield  {journal} {\bibinfo
  {journal} {Phys. Rev. A}\ }\textbf {\bibinfo {volume} {90}},\ \bibinfo
  {pages} {024304} (\bibinfo {year} {2014})}\BibitemShut {NoStop}%
\bibitem [{\citenamefont {Regula}\ \emph {et~al.}(2014)\citenamefont {Regula},
  \citenamefont {Di~Martino}, \citenamefont {Lee},\ and\ \citenamefont
  {Adesso}}]{PhysRevLett.113.110501}%
  \BibitemOpen
  \bibfield  {author} {\bibinfo {author} {\bibfnamefont {B.}~\bibnamefont
  {Regula}}, \bibinfo {author} {\bibfnamefont {S.}~\bibnamefont {Di~Martino}},
  \bibinfo {author} {\bibfnamefont {S.}~\bibnamefont {Lee}}, \ and\ \bibinfo
  {author} {\bibfnamefont {G.}~\bibnamefont {Adesso}},\ }\href {\doibase
  10.1103/PhysRevLett.113.110501} {\bibfield  {journal} {\bibinfo  {journal}
  {Phys. Rev. Lett.}\ }\textbf {\bibinfo {volume} {113}},\ \bibinfo {pages}
  {110501} (\bibinfo {year} {2014})}\BibitemShut {NoStop}%
\bibitem [{\citenamefont {Choi}\ and\ \citenamefont
  {Kim}(2015)}]{PhysRevA.92.042307}%
  \BibitemOpen
  \bibfield  {author} {\bibinfo {author} {\bibfnamefont {J.~H.}\ \bibnamefont
  {Choi}}\ and\ \bibinfo {author} {\bibfnamefont {J.~S.}\ \bibnamefont {Kim}},\
  }\href {\doibase 10.1103/PhysRevA.92.042307} {\bibfield  {journal} {\bibinfo
  {journal} {Phys. Rev. A}\ }\textbf {\bibinfo {volume} {92}},\ \bibinfo
  {pages} {042307} (\bibinfo {year} {2015})}\BibitemShut {NoStop}%
\bibitem [{\citenamefont {Karmakar}\ \emph {et~al.}(2016)\citenamefont
  {Karmakar}, \citenamefont {Sen}, \citenamefont {Bhar},\ and\ \citenamefont
  {Sarkar}}]{PhysRevA.93.012327}%
  \BibitemOpen
  \bibfield  {author} {\bibinfo {author} {\bibfnamefont {S.}~\bibnamefont
  {Karmakar}}, \bibinfo {author} {\bibfnamefont {A.}~\bibnamefont {Sen}},
  \bibinfo {author} {\bibfnamefont {A.}~\bibnamefont {Bhar}}, \ and\ \bibinfo
  {author} {\bibfnamefont {D.}~\bibnamefont {Sarkar}},\ }\href {\doibase
  10.1103/PhysRevA.93.012327} {\bibfield  {journal} {\bibinfo  {journal} {Phys.
  Rev. A}\ }\textbf {\bibinfo {volume} {93}},\ \bibinfo {pages} {012327}
  (\bibinfo {year} {2016})}\BibitemShut {NoStop}%
\bibitem [{\citenamefont {Adesso}\ and\ \citenamefont
  {Illuminati}(2007)}]{GaussianSM}%
  \BibitemOpen
  \bibfield  {author} {\bibinfo {author} {\bibfnamefont {G.}~\bibnamefont
  {Adesso}}\ and\ \bibinfo {author} {\bibfnamefont {F.}~\bibnamefont
  {Illuminati}},\ }\href {\doibase 10.1103/PhysRevLett.99.150501} {\bibfield
  {journal} {\bibinfo  {journal} {Phys. Rev. Lett.}\ }\textbf {\bibinfo
  {volume} {99}},\ \bibinfo {pages} {150501} (\bibinfo {year}
  {2007})}\BibitemShut {NoStop}%
\bibitem [{\citenamefont {Regula}\ \emph
  {et~al.}(2016{\natexlab{a}})\citenamefont {Regula}, \citenamefont
  {Osterloh},\ and\ \citenamefont {Adesso}}]{PhysRevA.93.052338}%
  \BibitemOpen
  \bibfield  {author} {\bibinfo {author} {\bibfnamefont {B.}~\bibnamefont
  {Regula}}, \bibinfo {author} {\bibfnamefont {A.}~\bibnamefont {Osterloh}}, \
  and\ \bibinfo {author} {\bibfnamefont {G.}~\bibnamefont {Adesso}},\ }\href
  {\doibase 10.1103/PhysRevA.93.052338} {\bibfield  {journal} {\bibinfo
  {journal} {Phys. Rev. A}\ }\textbf {\bibinfo {volume} {93}},\ \bibinfo
  {pages} {052338} (\bibinfo {year} {2016}{\natexlab{a}})}\BibitemShut
  {NoStop}%
\bibitem [{\citenamefont {Luo}\ and\ \citenamefont {Li}(2015)}]{LUO2015511}%
  \BibitemOpen
  \bibfield  {author} {\bibinfo {author} {\bibfnamefont {Y.}~\bibnamefont
  {Luo}}\ and\ \bibinfo {author} {\bibfnamefont {Y.}~\bibnamefont {Li}},\
  }\href {\doibase https://doi.org/10.1016/j.aop.2015.08.022} {\bibfield
  {journal} {\bibinfo  {journal} {Annals of Physics}\ }\textbf {\bibinfo
  {volume} {362}},\ \bibinfo {pages} {511} (\bibinfo {year}
  {2015})}\BibitemShut {NoStop}%
\bibitem [{\citenamefont {Verstraete}\ \emph {et~al.}(2002)\citenamefont
  {Verstraete}, \citenamefont {Dehaene}, \citenamefont {De~Moor},\ and\
  \citenamefont {Verschelde}}]{PhysRevA.65.052112}%
  \BibitemOpen
  \bibfield  {author} {\bibinfo {author} {\bibfnamefont {F.}~\bibnamefont
  {Verstraete}}, \bibinfo {author} {\bibfnamefont {J.}~\bibnamefont {Dehaene}},
  \bibinfo {author} {\bibfnamefont {B.}~\bibnamefont {De~Moor}}, \ and\
  \bibinfo {author} {\bibfnamefont {H.}~\bibnamefont {Verschelde}},\ }\href
  {\doibase 10.1103/PhysRevA.65.052112} {\bibfield  {journal} {\bibinfo
  {journal} {Phys. Rev. A}\ }\textbf {\bibinfo {volume} {65}},\ \bibinfo
  {pages} {052112} (\bibinfo {year} {2002})}\BibitemShut {NoStop}%
\bibitem [{\citenamefont {Regula}\ \emph
  {et~al.}(2016{\natexlab{b}})\citenamefont {Regula}, \citenamefont
  {Di~Martino}, \citenamefont {Lee},\ and\ \citenamefont
  {Adesso}}]{PhysRevLett.116.049902}%
  \BibitemOpen
  \bibfield  {author} {\bibinfo {author} {\bibfnamefont {B.}~\bibnamefont
  {Regula}}, \bibinfo {author} {\bibfnamefont {S.}~\bibnamefont {Di~Martino}},
  \bibinfo {author} {\bibfnamefont {S.}~\bibnamefont {Lee}}, \ and\ \bibinfo
  {author} {\bibfnamefont {G.}~\bibnamefont {Adesso}},\ }\href {\doibase
  10.1103/PhysRevLett.116.049902} {\bibfield  {journal} {\bibinfo  {journal}
  {Phys. Rev. Lett.}\ }\textbf {\bibinfo {volume} {116}},\ \bibinfo {pages}
  {049902} (\bibinfo {year} {2016}{\natexlab{b}})}\BibitemShut {NoStop}%
\bibitem [{\citenamefont {Chen}\ \emph {et~al.}(2005)\citenamefont {Chen},
  \citenamefont {Albeverio},\ and\ \citenamefont
  {Fei}}]{PhysRevLett.95.040504}%
  \BibitemOpen
  \bibfield  {author} {\bibinfo {author} {\bibfnamefont {K.}~\bibnamefont
  {Chen}}, \bibinfo {author} {\bibfnamefont {S.}~\bibnamefont {Albeverio}}, \
  and\ \bibinfo {author} {\bibfnamefont {S.-M.}\ \bibnamefont {Fei}},\ }\href
  {\doibase 10.1103/PhysRevLett.95.040504} {\bibfield  {journal} {\bibinfo
  {journal} {Phys. Rev. Lett.}\ }\textbf {\bibinfo {volume} {95}},\ \bibinfo
  {pages} {040504} (\bibinfo {year} {2005})}\BibitemShut {NoStop}%
\bibitem [{\citenamefont {Mintert}\ and\ \citenamefont
  {Buchleitner}(2007)}]{PhysRevLett.98.140505}%
  \BibitemOpen
  \bibfield  {author} {\bibinfo {author} {\bibfnamefont {F.}~\bibnamefont
  {Mintert}}\ and\ \bibinfo {author} {\bibfnamefont {A.}~\bibnamefont
  {Buchleitner}},\ }\href {\doibase 10.1103/PhysRevLett.98.140505} {\bibfield
  {journal} {\bibinfo  {journal} {Phys. Rev. Lett.}\ }\textbf {\bibinfo
  {volume} {98}},\ \bibinfo {pages} {140505} (\bibinfo {year}
  {2007})}\BibitemShut {NoStop}%
\bibitem [{\citenamefont {Zhang}\ \emph {et~al.}(2008)\citenamefont {Zhang},
  \citenamefont {Gong}, \citenamefont {Zhang},\ and\ \citenamefont
  {Guo}}]{PhysRevA.78.042308}%
  \BibitemOpen
  \bibfield  {author} {\bibinfo {author} {\bibfnamefont {C.-J.}\ \bibnamefont
  {Zhang}}, \bibinfo {author} {\bibfnamefont {Y.-X.}\ \bibnamefont {Gong}},
  \bibinfo {author} {\bibfnamefont {Y.-S.}\ \bibnamefont {Zhang}}, \ and\
  \bibinfo {author} {\bibfnamefont {G.-C.}\ \bibnamefont {Guo}},\ }\href
  {\doibase 10.1103/PhysRevA.78.042308} {\bibfield  {journal} {\bibinfo
  {journal} {Phys. Rev. A}\ }\textbf {\bibinfo {volume} {78}},\ \bibinfo
  {pages} {042308} (\bibinfo {year} {2008})}\BibitemShut {NoStop}%
\bibitem [{\citenamefont {Osborne}(2005)}]{PhysRevA.72.022309}%
  \BibitemOpen
  \bibfield  {author} {\bibinfo {author} {\bibfnamefont {T.~J.}\ \bibnamefont
  {Osborne}},\ }\href {\doibase 10.1103/PhysRevA.72.022309} {\bibfield
  {journal} {\bibinfo  {journal} {Phys. Rev. A}\ }\textbf {\bibinfo {volume}
  {72}},\ \bibinfo {pages} {022309} (\bibinfo {year} {2005})}\BibitemShut
  {NoStop}%
\bibitem [{\citenamefont {Lohmayer}\ \emph {et~al.}(2006)\citenamefont
  {Lohmayer}, \citenamefont {Osterloh}, \citenamefont {Siewert},\ and\
  \citenamefont {Uhlmann}}]{PhysRevLett.97.260502}%
  \BibitemOpen
  \bibfield  {author} {\bibinfo {author} {\bibfnamefont {R.}~\bibnamefont
  {Lohmayer}}, \bibinfo {author} {\bibfnamefont {A.}~\bibnamefont {Osterloh}},
  \bibinfo {author} {\bibfnamefont {J.}~\bibnamefont {Siewert}}, \ and\
  \bibinfo {author} {\bibfnamefont {A.}~\bibnamefont {Uhlmann}},\ }\href
  {\doibase 10.1103/PhysRevLett.97.260502} {\bibfield  {journal} {\bibinfo
  {journal} {Phys. Rev. Lett.}\ }\textbf {\bibinfo {volume} {97}},\ \bibinfo
  {pages} {260502} (\bibinfo {year} {2006})}\BibitemShut {NoStop}%
\bibitem [{\citenamefont {Osterloh}\ \emph {et~al.}(2008)\citenamefont
  {Osterloh}, \citenamefont {Siewert},\ and\ \citenamefont
  {Uhlmann}}]{PhysRevA.77.032310}%
  \BibitemOpen
  \bibfield  {author} {\bibinfo {author} {\bibfnamefont {A.}~\bibnamefont
  {Osterloh}}, \bibinfo {author} {\bibfnamefont {J.}~\bibnamefont {Siewert}}, \
  and\ \bibinfo {author} {\bibfnamefont {A.}~\bibnamefont {Uhlmann}},\ }\href
  {\doibase 10.1103/PhysRevA.77.032310} {\bibfield  {journal} {\bibinfo
  {journal} {Phys. Rev. A}\ }\textbf {\bibinfo {volume} {77}},\ \bibinfo
  {pages} {032310} (\bibinfo {year} {2008})}\BibitemShut {NoStop}%
\bibitem [{\citenamefont {Eltschka}\ and\ \citenamefont
  {Siewert}(2012)}]{PhysRevLett.108.020502}%
  \BibitemOpen
  \bibfield  {author} {\bibinfo {author} {\bibfnamefont {C.}~\bibnamefont
  {Eltschka}}\ and\ \bibinfo {author} {\bibfnamefont {J.}~\bibnamefont
  {Siewert}},\ }\href {\doibase 10.1103/PhysRevLett.108.020502} {\bibfield
  {journal} {\bibinfo  {journal} {Phys. Rev. Lett.}\ }\textbf {\bibinfo
  {volume} {108}},\ \bibinfo {pages} {020502} (\bibinfo {year}
  {2012})}\BibitemShut {NoStop}%
\bibitem [{\citenamefont {Eltschka}\ and\ \citenamefont
  {Siewert}(2014)}]{PhysRevA.89.022312}%
  \BibitemOpen
  \bibfield  {author} {\bibinfo {author} {\bibfnamefont {C.}~\bibnamefont
  {Eltschka}}\ and\ \bibinfo {author} {\bibfnamefont {J.}~\bibnamefont
  {Siewert}},\ }\href {\doibase 10.1103/PhysRevA.89.022312} {\bibfield
  {journal} {\bibinfo  {journal} {Phys. Rev. A}\ }\textbf {\bibinfo {volume}
  {89}},\ \bibinfo {pages} {022312} (\bibinfo {year} {2014})}\BibitemShut
  {NoStop}%
\bibitem [{\citenamefont {Siewert}\ and\ \citenamefont
  {Eltschka}(2012)}]{PhysRevLett.108.230502}%
  \BibitemOpen
  \bibfield  {author} {\bibinfo {author} {\bibfnamefont {J.}~\bibnamefont
  {Siewert}}\ and\ \bibinfo {author} {\bibfnamefont {C.}~\bibnamefont
  {Eltschka}},\ }\href {\doibase 10.1103/PhysRevLett.108.230502} {\bibfield
  {journal} {\bibinfo  {journal} {Phys. Rev. Lett.}\ }\textbf {\bibinfo
  {volume} {108}},\ \bibinfo {pages} {230502} (\bibinfo {year}
  {2012})}\BibitemShut {NoStop}%
\bibitem [{\citenamefont {Lancien}\ \emph {et~al.}(2016)\citenamefont
  {Lancien}, \citenamefont {Di~Martino}, \citenamefont {Huber}, \citenamefont
  {Piani}, \citenamefont {Adesso},\ and\ \citenamefont
  {Winter}}]{PhysRevLett.117.060501}%
  \BibitemOpen
  \bibfield  {author} {\bibinfo {author} {\bibfnamefont {C.}~\bibnamefont
  {Lancien}}, \bibinfo {author} {\bibfnamefont {S.}~\bibnamefont {Di~Martino}},
  \bibinfo {author} {\bibfnamefont {M.}~\bibnamefont {Huber}}, \bibinfo
  {author} {\bibfnamefont {M.}~\bibnamefont {Piani}}, \bibinfo {author}
  {\bibfnamefont {G.}~\bibnamefont {Adesso}}, \ and\ \bibinfo {author}
  {\bibfnamefont {A.}~\bibnamefont {Winter}},\ }\href {\doibase
  10.1103/PhysRevLett.117.060501} {\bibfield  {journal} {\bibinfo  {journal}
  {Phys. Rev. Lett.}\ }\textbf {\bibinfo {volume} {117}},\ \bibinfo {pages}
  {060501} (\bibinfo {year} {2016})}\BibitemShut {NoStop}%
\bibitem [{\citenamefont {Giampaolo}\ \emph {et~al.}(2011)\citenamefont
  {Giampaolo}, \citenamefont {Gualdi}, \citenamefont {Monras},\ and\
  \citenamefont {Illuminati}}]{PhysRevLett.107.260602}%
  \BibitemOpen
  \bibfield  {author} {\bibinfo {author} {\bibfnamefont {S.~M.}\ \bibnamefont
  {Giampaolo}}, \bibinfo {author} {\bibfnamefont {G.}~\bibnamefont {Gualdi}},
  \bibinfo {author} {\bibfnamefont {A.}~\bibnamefont {Monras}}, \ and\ \bibinfo
  {author} {\bibfnamefont {F.}~\bibnamefont {Illuminati}},\ }\href {\doibase
  10.1103/PhysRevLett.107.260602} {\bibfield  {journal} {\bibinfo  {journal}
  {Phys. Rev. Lett.}\ }\textbf {\bibinfo {volume} {107}},\ \bibinfo {pages}
  {260602} (\bibinfo {year} {2011})}\BibitemShut {NoStop}%
\bibitem [{\citenamefont {Panda}\ and\ \citenamefont
  {Benjamin}(2025)}]{PhysRevA.111.012420}%
  \BibitemOpen
  \bibfield  {author} {\bibinfo {author} {\bibfnamefont {D.~K.}\ \bibnamefont
  {Panda}}\ and\ \bibinfo {author} {\bibfnamefont {C.}~\bibnamefont
  {Benjamin}},\ }\href {\doibase 10.1103/PhysRevA.111.012420} {\bibfield
  {journal} {\bibinfo  {journal} {Phys. Rev. A}\ }\textbf {\bibinfo {volume}
  {111}},\ \bibinfo {pages} {012420} (\bibinfo {year} {2025})}\BibitemShut
  {NoStop}%
\bibitem [{\citenamefont {Zhao}\ \emph {et~al.}(2025)\citenamefont {Zhao},
  \citenamefont {Hou}, \citenamefont {He}, \citenamefont {Lo~Piparo},\ and\
  \citenamefont {Meng}}]{PhysRevA.111.042429}%
  \BibitemOpen
  \bibfield  {author} {\bibinfo {author} {\bibfnamefont {Y.}~\bibnamefont
  {Zhao}}, \bibinfo {author} {\bibfnamefont {J.}~\bibnamefont {Hou}}, \bibinfo
  {author} {\bibfnamefont {K.}~\bibnamefont {He}}, \bibinfo {author}
  {\bibfnamefont {N.}~\bibnamefont {Lo~Piparo}}, \ and\ \bibinfo {author}
  {\bibfnamefont {X.}~\bibnamefont {Meng}},\ }\href {\doibase
  10.1103/PhysRevA.111.042429} {\bibfield  {journal} {\bibinfo  {journal}
  {Phys. Rev. A}\ }\textbf {\bibinfo {volume} {111}},\ \bibinfo {pages}
  {042429} (\bibinfo {year} {2025})}\BibitemShut {NoStop}%
\end{thebibliography}

%

\end{document}